\documentclass[12pt]{article}

\advance\voffset by -0.25cm
\usepackage[utf8x]{inputenc}
\usepackage[round]{natbib}

\usepackage{amsmath,amssymb}
\usepackage{bbm}
\usepackage{graphicx}
\usepackage{longtable,dcolumn,booktabs}
\usepackage{textcomp}
\usepackage[font=small,labelfont=bf]{caption}

\def\calN{\mathcal{N}}
\def\calS{\mathcal{S}}
\def\calT{\mathcal{T}}
\def\calW{\mathcal{W}}

\newcommand{\E}{\ensuremath{\mathbf{E}}}
\newcommand{\Ensf}{\ensuremath{\mathbf{f}}}

\def\indicator#1{\mathbf{1}_{#1}}
\def\MD{{\mbox{\it MD}}}
\def\crps{{\mbox{CRPS}}}

\begin{document}

\title{Probabilistic Quantitative Precipitation Forecasting Using Ensemble Model Output Statistics}

\author{Michael Scheuerer}

\maketitle

\begin{abstract}
 Statistical post-processing of dynamical forecast ensembles is an essential component of weather forecasting. In this article, we present a post-processing method that generates full predictive probability distributions for precipitation accumulations based on ensemble model output statistics (EMOS).\\
 We model precipitation amounts by a generalized extreme value distribution that is left-censored at zero. This distribution permits modelling precipitation on the original scale without prior transformation of the data. A closed form expression for its continuous rank probability score can be derived and permits computationally efficient model fitting. We discuss an extension of our approach that incorporates further statistics characterizing the spatial variability of precipitation amounts in the vicinity of the location of interest.\\
 The proposed EMOS method is applied to daily 18-h forecasts of {6-h} accumulated precipitation over Germany in 2011 using the COSMO-DE ensemble prediction system operated by the German Meteorological Service. It yields calibrated and sharp predictive distributions and compares favourably with extended logistic regression and Bayesian model averaging which are state of the art approaches for precipitation post-processing. The incorporation of neighbourhood information further improves predictive performance and turns out to be a useful strategy to account for displacement errors of the  dynamical forecasts in a probabilistic forecasting framework.
\end{abstract}

\section{Introduction}
\label{intro}
In recent years, weather prediction has seen a culture change towards probabilistic forecasting. In order to represent forecast uncertainties, ensembles of dynamical forecasts are generated with members corresponding to model integrations that differ in the initial conditions and/or the numerical representation of the atmosphere \citep{Palmer2002}. These are the main sources of uncertainty, but the different ensemble members still share certain structural model deficiencies and usually fail to represent the full uncertainty that comes with numerical weather prediction. Statistical post-processing has therefore become an integral part of any ensemble prediction system, aiming to remove systematic biases and to achieve appropriate representation of the forecast uncertainty \citep{GneitingRaftery2005}.

Among the various approaches to statistical post-processing, methods that transform the ensemble forecasts into a full predictive cumulative distribution function (CDF) function are particularly convenient, because all kinds of probabilistic statements (prediction intervals, probabilities of threshold exceedance, etc.) can be derived from this CDF in a consistent way. Finding a suitable probabilistic model is more challenging for precipitation than for most other weather variables because the associated uncertainty calls for rather special distributions with the following features:
\begin{itemize}
 \item they must be non-negative
 \item they may be equal to zero with positive probability
 \item their non-zero component has positive skew
\end{itemize}
Several authors \citep{HamillColucci1997, Sloughter&2007, Wilks2009} report an improved fit if their models are fitted to powers (typically between $0.25$ and $0.5$) of forecasts and observations, which suggests that the predictive distribution for precipitation amounts on the original scale has a heavy right tail. An inconvenient side effect of such power transformations is, however, that they distort the original scale, accentuating ensemble spread at low precipitation levels and attenuating it at higher ones. To avoid this effect, we will use a family of distributions that can be used directly with the untransformed data.

We adopt the paradigm formulated by \citet{BroeckerSmith2008}, that post-processing should make optimal use of the {\em information} contained in the ensemble without relying on any assumption about ensemble forecasts being draws from some unknown distributions. Our aim is to develop a model that is of comparable conceptional simplicity and computational efficiency as the extended
logistic regression approach by \citet{Wilks2009}, but in addition permits the inclusion of {\em uncertainty information from the ensemble} in a natural and intuitive way. To achieve this, we adapt the non-homogeneous Gaussian regression approach by \citet{Gneiting&2005} so as to respect the peculiarities of precipitation.
In Sec.~\ref{sec:2} we introduce an adaptation of the generalized extreme value (GEV) distribution family, and argue that it presents, for reasonable choices of the shape parameter, an ideal candidate for predictive distributions for quantitative precipitation. We further discuss suitable predictor variables for the GEV location and scale parameter that convey the relevant information of the ensemble. A closed form expression for the continuous rank probability score (CRPS) of the adapted GEV is provided and will be used for model fitting. In Sec.~\ref{sec:3}, the issue of displacement errors of dynamical precipitation forecasts is addressed. These errors are another peculiarity of precipitation and a further source of uncertainty. This uncertainty can be taken into account by our EMOS method by considering, for each location of interest, the dynamical forecasts in a larger neighbourhood of this location, and condensing this neighbourhood information into a further predictor. In Section {\ref{sec:4} we apply 
our method to daily 18-h forecasts of 6-h accumulated precipitation over Germany in 2011 generated by the COSMO-DE ensemble prediction system from the German Meteorological Service. A discussion of possible further extensions to our approach is subject of Sec.~\ref{sec:5}.

\section{A distribution family for quantitative precipitation}
\label{sec:2}

Consider the CDF of the generalized extreme value (GEV) distribution
\[
 G(y) := \begin{cases}
         \exp\left(-\left(1+\xi\left(\frac{y-\mu}{\sigma}\right)\right)^{-1/\xi}\right), & \xi \neq 0 \\
         \exp\left(-\exp\left(-\frac{y-\mu}{\sigma}\right)\right), & \xi = 0 \\
        \end{cases}
\]
which is a family of continuous distributions with parameters $\mu,\sigma$ and $\xi$ that characterize location, scale and shape of the GEV. For $\xi<0, y>\mu-\frac{\sigma}{\xi}$ one defines $G(y):=1$, for $\xi>0, y<\mu-\frac{\sigma}{\xi}$ one defines $G(y):=0$. In this paper we will always assume $\xi\in(-0.278,1)$, because for this range of values, the GEV has positive skew and its mean exists and is equal to
\[
 m = \begin{cases}
      \mu+\sigma\frac{\Gamma(1-\xi)-1}{\xi}, &  \xi\neq0 \\
      \mu+\sigma\gamma, & \xi=0      
     \end{cases}
\]
where $\Gamma$ denotes the gamma function and $\gamma\approx0.5772$ is the Euler-Mascheroni constant. To obtain a suitable model for precipitation amounts, we consider the GEV to be left-censored at zero, i.e.\ all mass below zero is assigned to exactly zero. The predictive CDF then becomes
\[
 \tilde{G}(y) := \begin{cases}
         G(y) & \mbox{ for } \; y \geq 0 \\
         \;0 & \mbox{ for } \; y < 0
        \end{cases}
\]
This distribution is non-negative and exactly zero with positive probability if either $\xi\leq0$ or $\xi>0$ and $\mu<\frac{\sigma}{\xi}$. For $\xi>0$ it has a heavy right tail, and so it accommodates all of the important features of a predictive distribution for quantitative precipitation forecasting mentioned above, and can be used directly with the untransformed data. To illustrate how the left-censored GEV distributions look in practice, Fig.~\ref{Fig:GEVdensities} shows the predictive density of 6-h precipitation accumulations on 11 June 2011 at various locations in Germany. Details about the data and the forecast ensemble are given later in Sec.~\ref{sec:4}. 

\begin{figure*}
 \centering
 \includegraphics[width=\textwidth]{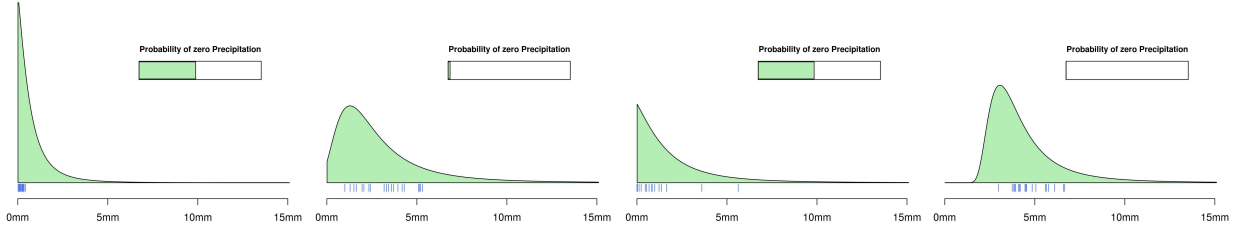}
 \caption{Predictive distributions of precipitation accumulations on 11 June 2011 between 1200 UTC and 1800 UTC at Stuttgart airport, Munich airport, Cologne/Bonn airport, and Munich city (from left to right). The short blue lines below the densities represent the ensemble member forecasts.}
 \label{Fig:GEVdensities}
\end{figure*}

\subsection*{Linking the parameters to ensemble model output statistics}

The next step after specifying a suitable family of predictive distributions is to link its parameters to suitable predictor variables. Although $m$ is no longer the mean of the left-censored GEV, it is a more suitable location parameter for our purposes than $\mu$, since it interacts more naturally with $\sigma$. Indeed, if for fixed $\mu$ the scale parameter $\sigma$ is increased, the whole mass of the (left-censored) GEV distribution is shifted to the right. If $m$ is kept fixed instead, then an increase of $\sigma$ spreads the mass of the distribution more symmetrically to both sides, which corresponds much better to the idea of increased uncertainty.
Now let $f_{s1},\ldots,f_{sK}$ be the ensemble member forecasts of precipitation amounts at location $s$. Their information will be condensed into the following statistics:
\begin{itemize}
 \item $\overline{\Ensf}_s:=\frac{1}{K}\sum_{k=1}^K f_{sk}$ \ (ensemble mean)
 \vspace{2mm}
 \item $\overline{\indicator{\{\Ensf_s=0\}}}:=\frac{1}{K}\sum_{k=1}^K \indicator{\{f_{sk}=0\}}$\ (fraction of zero precipitation members)
 \vspace{2mm}
 \item $\MD(\Ensf_s):=\frac{1}{K^2}\sum_{k,k^\prime=1}^K |f_{sk}-f_{sk^\prime}|$\ (ensemble mean difference)
\end{itemize}
While $\overline{\Ensf}_s$ is a standard predictor for the location parameter and $\overline{\indicator{\{\Ensf_s=0\}}}$ (or variants of it) can provide additional information about whether precipitation occurs or not \citep{Sloughter&2007, BentzienFriederichs2012}, $\MD(\Ensf_s)$ has not been employed in this context so far. It has certain properties, however, that make it an appealing dispersion measure for precipitation forecasts \citetext{see also \citealp{Yitzhaki2003}}:
\begin{itemize}
 \item it is more robust than the standard deviation because it uses absolute rather than squared differences
 \item unlike the interquartile range or the median absolute deviation it is sensitive to all ensemble forecasts
\end{itemize}
We let the parameters $m_s$ and $\sigma_s$ depend on the ensemble forecasts via
\[
 m_s = \alpha_0 + \alpha_1\cdot\overline{\Ensf}_s + \alpha_2\cdot\overline{\indicator{\{\Ensf_s=0\}}}, \qquad
 \sigma_s = \beta_0 + \beta_1\cdot\MD(\Ensf_s).
\]

\subsection*{Model fitting}

Given the ensemble member forecasts for some location, a predictive CDF for this location can be issued once suitable parameters $\alpha_0,\alpha_1,\alpha_2,\beta_0,\beta_1$ and $\xi$ have been selected. This is done based on a training set of forecasts-observations pairs. For each day, we let this training set consist of data from the preceding $n$ days and all rain-gauge locations within our domain of interest.

Proper scoring rules \citep{GneitingRaftery2007} are a standard tool for quantitative assessment of the quality of probabilistic forecasts. A scoring rule $s(F,y)$ assigns a numerical score to each pair $(F,y)$, where $F$ is the predictive CDF and $y$ is the verifying observation. If negatively oriented scores are considered, a lower score indicates a better probabilistic forecast, where ``better'' refers to both calibration and sharpness. These two properties should be the goal of probabilistic forecasting \citep{GneitingBalabdaouiRaftery2007}, and so it is natural to use strictly proper scoring rules as loss functions for training algorithms. With $\calT$ denoting the set of training days, $\calS$ the set of training sites, and $\#\calS$ its cardinality, the parameters are chosen such that the empirical score 
\begin{equation}\label{Eq:EmpiricalScore}
 \frac{1}{n\cdot\#\calS}\; \sum_{t\in\calT}\sum_{s\in\calS} S(F_{st},y_{st})
\end{equation}
is minimized. When it comes to choosing a specific scoring rule that is adequate in the present framework, the following aspects should be taken into account:
\begin{itemize}
 \item even state of the art high-resolution limited-area mesoscale NWP models such as the COSMO-DE-EPS frequently issue forecasts that are - at least when considered gridpoint by gridpoint - quite off the mark. The chosen scoring rule should therefore be reasonably robust and should not be corrupted by a few ``bad forecasts'' \citetext{see also the discussion in Section 5 of \citealp{BroeckerSmith2008}};
 \item the chosen scoring rule should be able to deal with the fact that the predictive distribution for precipitation has both a discrete and a continuous component. 
\end{itemize}
Both of these aspects make a strong case for the continuous ranked probability score
\begin{equation}\label{Eq:CRPS}
 \crps(F,y) = \int_{-\infty}^\infty \big(F(t)-{\bf 1}_{[y,\infty)}(t)\big)^2 dt
\end{equation}
\citep{Hersbach2000, GneitingRaftery2007}. A training algorithm based on minimum CRPS estimation typically requires many evaluations of \eqref{Eq:EmpiricalScore} which can be computationally intractable if the integral in \eqref{Eq:CRPS} has to be calculated numerically. However, in a recent paper \cite{FriederichsThorarinsdottir2012} derived a closed form expression for $\crps(G,y)$, and it is straightforward to generalize their calculations to the case of a left-censored GEV. For $\xi\neq0$ one obtains
\begin{eqnarray}
 \crps(\tilde{G}_{\xi\neq0},y) & = & (\mu-y)(1-2p_y) + \mu\,p_0^2 -\,2\,\frac{\sigma}{\xi}\, \Big[ 1-p_y -\Gamma_l(1-\xi,-\log p_y) \Big] \label{EqCRPScGEV} \\[1mm]
 & &  +\,\frac{\sigma}{\xi}\, \Big[ 1-p_0^2 - 2^\xi\,\Gamma_l(1-\xi,-2\log p_0) \Big] \nonumber
\end{eqnarray}
where
\[
 p_0 := G(0), \quad p_y := G(y),
\]
and $\Gamma_l$ denotes the lower incomplete gamma function.
A closed form expression for $\crps(\tilde{G}_{\xi=0},y)$ can be derived as well, but for numerical reasons it is preferable to use, for $\xi\in(-\epsilon,\epsilon)$ with $\epsilon$ reasonably small, the approximation
\[
 \crps(\tilde{G}_{\xi=0},y) = \tfrac{(\epsilon-\xi)}{2\epsilon}\:\crps(\tilde{G}_{\xi=-\epsilon},y) + \tfrac{(\epsilon+\xi)}{2\epsilon}\:\crps(\tilde{G}_{\xi=\epsilon},y).
\]
where the two scores on the right hand side are calculated according to \eqref{EqCRPScGEV}. Since $\Gamma_l$ is the product of a gamma function and the CDF of the gamma distribution, all of the above terms can be calculated with standard statistical software. Owing to this closed form, minimum CRPS estimation is computationally efficient and feasible even for large training sets.

\subsection*{Choice of the training period and regularization}

In our experiments with the COSMO-DE-EPS forecasts in Sec.~\ref{sec:4} we use a rolling training period of $n=30$ days, i.e.\ at each verification day we fit our model to forecast-observation pairs from the preceding $30$ days (as we consider lead times of 18h and less, this data would be available at the time where the new forecast is issued). One therefore obtains a different post-processing model for each verification day, which allows one to adapt to seasonal changes. The choice of the training period implies a bias/variance tradeoff. Longer periods imply more training data and lead to more stable parameter estimates, while shorter periods permit a more flexible reaction to seasonal changes. With our choice of $n=30$ days we ensure there are a sufficient number of wet days, but nevertheless some of the parameter estimates show unrealistically strong fluctuations over time (see Fig.~\ref{Fig:TempCycleLoc}). This suggests that a longer training period could be favourable, but instead of tuning $n$ to our 
specific data set, we investigate a different strategy to stabilize the estimates.

\begin{figure*}
 \centering
 \includegraphics[width=1\textwidth]{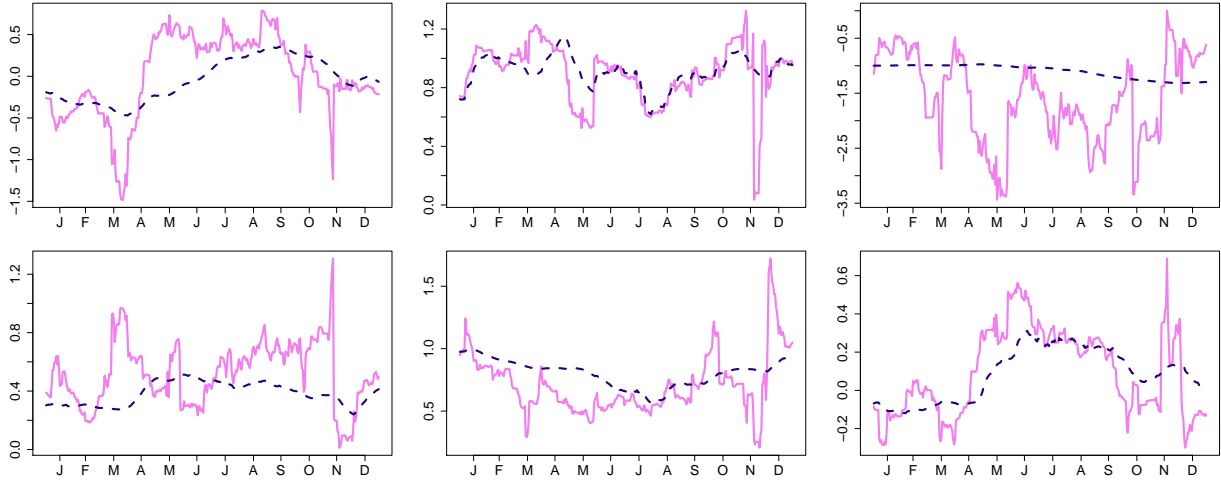}
 \caption{Temporal evolution of the location parameters $\alpha_0$ (top left), $\alpha_1$ (top), and $\alpha_2$ (top right), the scale parameters $\beta_0$ (bottom left) and $\beta_1$ (bottom), and the shape parameter $\xi$ (bottom right) obtained with standard CRPS minimization (solid lines) and early stopping (dashed lines).}
 \label{Fig:TempCycleLoc}
\end{figure*}

Generally, we expect the post-processing parameters to vary over time, but in such a way that changes are incremental. Because of this, it certainly makes sense to use the parameters chosen for one day as starting values for the training algorithm on the next day. Given the new training data, gradient-based optimization routines adjust these initial parameters typically by first taking a step in the direction of the steepest descent of the objective function and then proceeding with further refinements until the reduction in the objective is within a specified tolerance. To avoid overfitting, we use a regularization strategy which is referred to as ``early stopping'' in the machine learning and statistics literature. Instead of iterating the optimization routine to convergence, one stops after a few iterations. The rationale behind this strategy is that the important adjustments to the parameters are made during the first steps, while further adjustments often improve the fit to unimportant or even random 
features in the data only. Following this idea, and assuming that the parameters of the preceding day are excellent starting values, we stop numerical optimization after just one iteration of the quasi-Newton Broyden-Fletcher-Goldfarb-Shanno (BFGS) algorithm implemented in R \citep{R2010}. Note that this early stopping implies that the CRPS minimization based on the {\em training data} is incomplete. It turns out, however, that the verification results obtained with the regularized parameter estimates are even better than those obtained without regularization. More importantly, the temporal evolution of the different parameters becomes much smoother and hence more realistic (see Fig.~\ref{Fig:TempCycleLoc}). On the first verification day, where we need to guess a set of initial parameters, we let the BFGS algorithm perform ten iterations to permit it get away from the starting values. We take these to be $(\alpha_0,\alpha_1,\alpha_2)=(0,1,-1)$, optimistically suggesting that $m_s$ closely follows the 
ensemble mean with no need for and intercept and a assuming a moderate negative contribution of $\overline{\indicator{\{\Ensf_s=0\}}}$. Likewise we take $(\beta_0,\beta_1)=(0.1,1)$, this time allowing for a small intercept to ensure the variance of $\tilde{G}$ is positive. We have no intuition about the shape parameter and start with $\xi=0$. In order to make sure that $\beta_0>0, \beta_1\geq0$ and $\xi\in(-0.278,1)$ on all verification days, we alter the objective function \eqref{Eq:EmpiricalScore} so as to return a high enough value (we take twice the value of the last ``admissible'' evaluation of the objective) whenever some parameter violates the above conditions. In our experimental runs, however, it never occurred that the BFGS algorithm attempted to evaluate the objective function with a parameter set outside the admissible range.

Fig.~\ref{Fig:TempCycleLoc} shows the temporal evolution of all parameters over our 365-day verification period. Apparently, some of the parameters are much more prone to overfitting than others. The parameter $\alpha_1$, for example, which relates the ensemble mean to the location parameter $m$, is certainly the most influential of all parameters, and is fitted in a more or less stable way even if the training algorithm is iterated to convergence. The fraction of zero precipitation members, on the contrary, while providing additional information for distinguishing wet and dry regimes, has far less impact on the CRPS. As a consequence, the estimates of $\alpha_2$ are much more volatile, but can be stabilized by the proposed strategy. All remaining parameters are somewhere in between, and early stopping smoothes their temporal evolution while maintaining their seasonal cycles. Overall we conclude that early stopping yields - in addition to the computational speed-up - the benefit of avoiding overfitting of 
the post-processing parameters, and can thus even improve predictive performance. A similar conclusion was noted in a different context by \cite{Hamill2007}, who criticized the EM algorithm used in the Bayesian model averaging technique \citep{Raftery&2005} for assigning radically unequal weights to different ensemble members, and explained this as well by overfitting.

\section{Addressing displacement errors by using neighbourhood information}
\label{sec:3}

The method presented in Sec.~\ref{sec:2} can be used right away to turn ensemble forecasts into a predictive distribution, but there is another peculiarity about precipitation that a good post-processing method should take into account: the issue of displacement errors. Precipitation results from complex dynamical and microphysical processes, and precipitation forecasts will strongly be affected by uncertainties in the forecasts of many other variables in the NWP model. Underestimating the velocity of an approaching front, for example, will entail misallocation of forecast precipitation amounts at locations around the true and the forecast front position at the considered time. Even more challenging is the situation where precipitation results from locally forced convection, and predictability is limited by the short time scales of these processes. In both cases, the quantitative precipitation forecasts at specific locations by the NWP model may be quite far off, even if the dynamics are captured well 
qualitatively - the ``correct'' precipitation amount may simply be displaced by a certain distance. To address such displacement errors we consider ensemble forecasts in a neighbourhood $\calN(s)$ of radius $r$ around the location $s$ of interest. More specifically, we pass from $f_{s1},\ldots,f_{sK}$ to {\em weighted averages} of the forecasts at all gridpoints $x$ within $\calN(s)$:
\[
 f_{\calN(s),k} := \sum_{x\in\calN(s)}w_x^{(s)}f_{xk}, \quad k = 1,\ldots,K.
\]
Rather than using equal weights $w_x^{(s)}$ within the neighbourhood, we let them decay smoothly with distance from $s$
\[
 w_x^{(s)}\sim \max\left\{1-\left(\frac{dist(x,s)}{r}\right)^2,0\right\}
\]
and rescale them such that $\sum_{x\in\calN(s)}w_x^{(s)} = 1$. The rationale behind this choice is as follows: on the one hand we think that due to displacement, the ensemble forecasts at some point $x$ near $s$ might be closer to the truth than those at $s$ itself. On the other hand, we still think that the further we move away from $s$, the less likely it is, in absence of further information about a possible displacement, that the respective forecasts are more appropriate. In principle, this idea could be generalized and elliptic neighbourhoods could be used in order to account for anisotropy in location uncertainty. Or, if systematic displacement errors are observed, this could be corrected by shifting the centre of $\calN(s)$. Both extensions are likely to yield further improvement, but linking shapes and shifts of the neighbourhood to suitable covariates is all but straightforward, and for simplicity we shall stay for now with radially symmetric, unshifted neighbourhoods.
The original predictors from Sec.~\ref{sec:2} become
\begin{itemize}
 \item $\overline{\Ensf}_{\calN(s)}:=\frac{1}{K}\sum_{k=1}^K f_{\calN(s),k}\quad$ \ (mean weighted neighbourhood average)
 \vspace{2mm}
 \item $\overline{\indicator{\calN(s),\{\Ensf_x=0\}}}:=\frac{1}{K}\sum_{k=1}^K \: \sum_{x\in\calN(s)}w_x^{(s)} \:\indicator{\{f_{xk}=0\}}$\\[1mm]
  (weighted fraction of zero precipitation members)
 \vspace{2mm}
 \item $\MD\big(\Ensf_{\calN(s)}\big):=\frac{1}{K^2}\sum_{k,k^\prime=1}^K \big|f_{\calN(s),k}-f_{\calN(s),k^\prime}\big|$\\[1mm]
  (mean diff.\ of weighted neighbourhood averages)
\end{itemize}
Their re-definition so far essentially replaces the original forecasts by smoothed forecasts. For the third predictor this can also be a drawback: it still carries information about forecast uncertainty due to uncertain initial conditions and model physics, but displacement uncertainty represented by the ensemble may even by partly smoothed out. We therefore add a further predictor
\[
 \overline{\MD_{\calN(s)}(\Ensf_x)}:=\frac{1}{K}\sum_{k=1}^K \sum_{x,x^\prime\in\calN(s)} w_x^{(s)}w_{x^\prime}^{(s)} \: \big|f_{xk}-f_{x^\prime k}\big|
\]
(mean neighbourhood weighted mean differences) which addresses uncertainty due to spatial variability of precipitation forecasts, but averages over the different ensemble members. The calculation of the double sum can be computationally expensive for large neighbourhoods, but by comparing equations (19) and (20) in \citet{Hersbach2000} one obtains
\begin{equation}\label{Eq:HersbachRepresentation}
 \sum_{x,x^\prime\in\calN(s)} w_x^{(s)}w_{x^\prime}^{(s)} \big|f_{xk}-f_{x^\prime k}\big| = 2 \sum_{i=1}^{N-1} \calW^{(i)} \big(1-\calW^{(i)}\big) \big(f_{x_{i+1}k}-f_{x_ik}\big)
\end{equation}
where $x_1,\ldots,x_N$ is, for any fixed $k$, an enumeration of all gridpoints in $\calN(s)$ such that $f_{x_1k}\leq\ldots\leq f_{x_Nk}$ and $\calW^{(i)}=\sum_{j=1}^i w_{x_j}^{(s)}$. The computational costs for sorting are negligible if efficient sorting algorithms such as quicksort or heapsort \citetext{see \citealp{Press&1989}} are used, and so passing from a double to a single sum can reduce computational time considerably. The location and scale parameters $m$ and $\sigma$ of our left-censored GEV are linked to the above statistics via
\begin{eqnarray*}
 m & = & \alpha_0 + \alpha_1\cdot\overline{\Ensf}_{\calN(s)} + \alpha_2\cdot\overline{\indicator{\calN(s),\{\Ensf_x=0\}}} \\
 \sigma & = & \beta_0 + \beta_1\cdot\MD\big(\Ensf_{\calN(s)}\big) + \beta_2\cdot\overline{\MD_{\calN(s)}(\Ensf_x)}
\end{eqnarray*}
As before, the parameters $\alpha_0,\alpha_1,\alpha_2,\beta_0,\beta_1,\beta_2$ and $\xi$ are fitted to training data via CRPS minimisation.

\begin{figure*}
 \centering
 \includegraphics[width=1\textwidth]{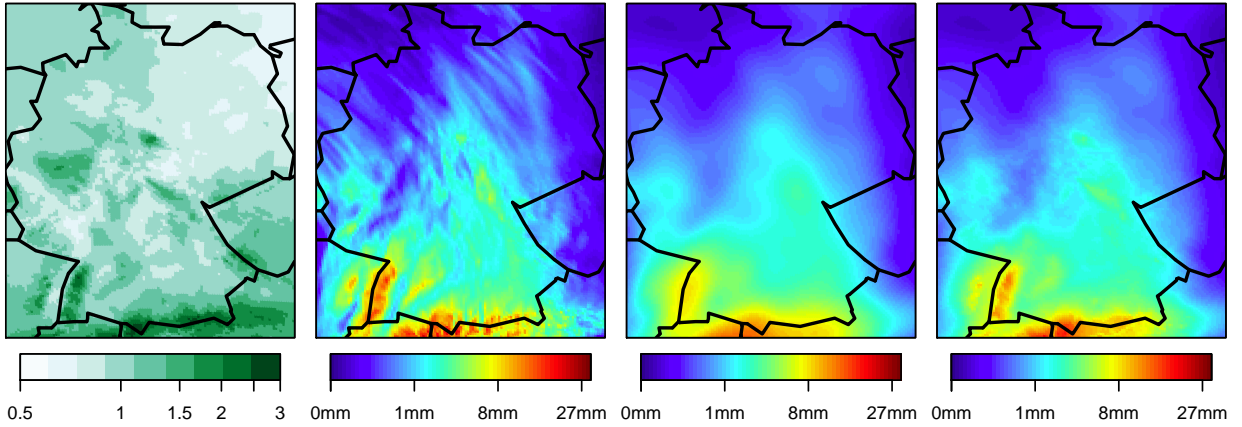}
 \caption{Standardization factor $c_s$ (left), ensemble mean of predicted precipitation for 12 December 2011 between 1200 UTC and 1800 UTC (middle left), and mean weighted 60 km-neighbourhood averages calculated on the original (middle right) and on the standardized (right) scale.}
 \label{Fig:Neighbourhood-Correction}
\end{figure*}

The choice of the neighbourhood radius $r$ certainly depends on both the forecast ensemble and the topography of the considered domain. Large neighbourhoods can make allowance for large displacement errors, but entail a blurring of precipitation processes thus reducing the spatial resolution of forecasts. One particular aspect of this is that orographically induced precipitation will be spread out to locations which are close-by but have an entirely different topography. At least this aspect can be accounted for by ``standardizing'' predicted precipitation amounts to make them independent of the orography, calculating the above predictor variables with the standardized forecasts, and finally transforming the resulting predictors back to the original, orography-dependent scale. As an indicator for orography-related spatial variations in precipitation patterns we define $c_s$ as the {\em mean annual precipitation amount} at gridpoint $s$ divided by the median of all of these values over the considered domain. 
We then replace $f_{xk}$ in the definitions of $\Ensf_{\calN(s)}$ and $\overline{\MD_{\calN(s)}(\Ensf_x)}$ by the standardized forecasts $f_{xk}^c:=f_{xk}/c_x$ and use
\[
 c_s\cdot\overline{\Ensf^c}_{\calN(s)}, \quad c_s\cdot\MD\big(\Ensf_{\calN(s)}^c\big), \quad\mbox{and}\quad c_s\cdot\overline{\MD_{\calN(s)}(\Ensf_x^c)}
\]
as predictors for the parameters $m$ and $\sigma$ (the remaining statistic, $\overline{\indicator{\calN(s),\{\Ensf_x=0\}}}$, is let unchanged). To calculate the standardization factor $c_s$ we used measurements of mean annual precipitation at rain gauges in Germany, Austria and Switzerland from the years 2005 to 2010. We fitted a 3D spatial Gaussian process model to these observations and interpolated them to the forecast grid (2.8 km horizontal resolution) via Kriging \citetext{cf.~\citealp{chiles-delfiner}}, a technique for spatial interpolation similar to what is referred to as ``objective analysis'' in the meteorological literature. The fitted 3D model describes how correlations between observations decay both horizontally and vertically. It thus provides a measure of similarity between different sites which is the starting point of the Kriging interpolation scheme. As the topography of Germany gets increasingly complex as one moves from north to south (starting with coastal areas, followed by the North 
German Plain, then by an area with various low mountain ranges, and finally the alpine foothills), additional data from the southern neighbours, Austria and Switzerland, was required to ensure one obtains reasonable interpolates in the alpine regions.

Figure \ref{Fig:Neighbourhood-Correction} illustrates the difference between the climatology-corrected and the uncorrected neighbourhood averaging scheme with precipitation forecasts over Germany. The standardization factor $c_s$ is shown in first plot. The second plot depicts the ensemble mean of predicted precipitation on a day where elevated precipitation levels are forecast in particular over several alpine regions, Black Forest (south-west) and, to a lesser extent, also some other mid-range mountains. If mean weighted neighbourhood averages with a neighbourhood radius of 60 km are considered instead of the ensemble mean, orographic effects are largely smoothed out (third plot). The corresponding statistics with the same neighbourhood size but climatology-correction account for displacement errors while keeping high precipitation levels linked to mountainous topography (fourth plot).

\section{Post-processing of COSMO-DE-EPS forecasts of precipitation accumulations over Germany in 2011}
\label{sec:4}

\subsection*{Forecast and observational datasets used}

We test the EMOS post-processing method introduced in Sections \ref{sec:2} and \ref{sec:3} with forecasts of 6-h accumulated precipitation from the COSMO-DE-EPS and rain gauge observations from about 1130 SYNOP stations in Germany. COSMO-DE-EPS is a multi-analysis and multi-physics ensemble prediction system based on the high-resolution numerical weather prediction (NWP) model COSMO-DE \citep{Baldauf&2011}, a configuration of the COSMO model with a horizontal grid size of 2.8 km operated by the German Meteorological Service (DWD). It is in pre-operational phase since 9 December 2010, covers the area of Germany and produces forecasts with lead times up to 21 hours. The current setup of the lateral boundary conditions uses forecasts from different global models, while different configurations of the COSMO-DE model are used for the variation of model physics. Deep convection is simulated explicitly, and the perturbations of the model physics are tailored to represent uncertainties in quantitative precipitation 
forecasts \citetext{for further details see \citealp{Gebhardt&2011}}. Preliminary studies show that ensemble precipitation forecasts by COSMO-DE-EPS are just slightly underdispersive and have a moderate tendency to overforecast precipitation amounts. Despite their good performance we think that probabilistic quantitative precipitation forecasts can be further improved by post-processing, especially during the summer months where precipitation is often due to locally forced convection with small convective cells whose highly random appearance can hardly be represented by only 20 ensemble members. We consider daily forecasts of precipitation amounts between 1200 UTC and 1800 UTC by the COSMO-DE-EPS model run initialized at 0000 UTC. The whole year 2011 will be used as a verification period, whence the respective training periods are a bit shorter than 30 days during the first few days.

\subsection*{Forecast verification techniques}

As a measure of predictive performance we use proper scoring rules \citep{JolliffeStephenson2003, Wilks2006, GneitingRaftery2007} which address both calibration and sharpness of the probabilistic forecasts simultaneously. In addition to the CRPS, introduced in Sec.~\ref{sec:2}, we consider Brier scores for the binary event $o$ that the observed precipitation amount $y$ exceeds a certain threshold $t$. If $p=1-F(t)$ denotes the predicted probability of this event, the Brier score is defined as
\[
 BS(p,o) = \big(o-p\big)^2 = \big(F(t)-{\bf 1}_{[y,\infty)}(t)\big)^2
\]
Brier scores are useful to check, for example, if probabilistic forecasts for high precipitation levels are reliable and issued with reasonable resolution, whereas the CRPS, being an integral over the BS at all thresholds, measures the overall performance. Since the frequency of observed 1's rapidly declines with increasing thresholds, the corresponding values of the BS have entirely different magnitudes. To facilitate a comparison, it is convenient to consider the Brier skill score
\[
 BSS = 1 - \frac{BS}{BS_{ref}}
\]
which is positively oriented and can be interpreted as the improvement over a reference forecast. In our study the reference forecast at each site $s$ will be the fraction $\frac{1}{K}\sum_{k=1}^K \indicator{\{f_{sk}>t\}}$ of ensemble forecasts above the threshold. In the same way, we will employ the continuous ranked probability skill score (CRPSS) which is defined analogously to the BSS, and uses the CRPS of the raw ensemble forecasts
\begin{equation}\label{Eq:CRPS-Ensemble}
 \crps(F_{ens},y) = \frac{1}{K}\sum_{k=1}^K |f_{k}-y| - \frac{2}{K^2}\sum_{k,k^\prime=1}^K|f_{k}-f_{k^\prime}|,
\end{equation}
where $F_{ens}=\frac{1}{K}\sum_{k=1}^K \indicator{[f_k,\infty)}(t)$, as a reference score. Expression \eqref{Eq:CRPS-Ensemble} is obtained from the alternative representation $\crps(F,y)=\E_F|X-y|-\frac{1}{2}\E_F|X-X^\prime|$ \citetext{see \citealp{GneitingRaftery2007}} by plugging in the empirical CDF of the ensemble forecasts.

\begin{table*}\footnotesize
 \caption{Brier and CRP skill scores for the different post-processing methods.}
 \centering
 \label{Tab:SkillScores}
 \begin{tabular}{lccccc}
  \toprule
   & BSS (0 mm) & BSS (5 mm) & BSS (10 mm) & BSS (15 mm) & CRPSS \\
  \midrule
  extended LR  & 0.079 & 0.044 & 0.048 & 0.044 & 0.052 \\
  BMA          & 0.042 & 0.008 & 0.022 & 0.031 & 0.024 \\
  EMOS         & 0.055 & 0.046 & 0.057 & 0.040 & 0.054 \\
  \bottomrule
 \end{tabular}
\end{table*}

In order to check whether possible differences in skill are significant we use statistical tests for equal performance. We will focus on the CRPS (which measures overall performance) and provide p-values obtained from paired t-tests. Following the recommendations of \citet{Hamill1999}, tests are performed using differences in daily sums of CRPS rather than differences in daily CRPSSs. The aggregation of scores to daily sums accounts for spatial dependence while scores are considered independent from one day to the next, an assumption supported by \citet{Hamill1999}, Table 3, and the fact that the accumulation periods considered here are 18h apart.

To assess the calibration of the probability over threshold forecasts by the raw ensemble on the one hand and our EMOS approach on the other hand, we use reliability diagrams \citetext{e.g.~\citealp{Wilks2006}, Ch.~7} enhanced with uncertainty information. The forecast probabilities are divided in 11 categories defined as follows:
\[
 B_1:=[0,0.05),B_2:=[0.05,0.15),\ldots,B_{11}:=[0.95,1.0].
\]
A forecast is reliable if the relative frequency of the event $o_i=1$, when computed over all instances $i$ for which $p_i$ falls into the interval $B_k$, is equal to the mean $\bar{\pi}_k$ of $p_i$ over that interval \citep{BroeckerSmith2007}. Especially for the first bin and high threshold values, $\bar{\pi}_k$ can be quite different from the arithmetic centre of $B_k$, and so we follow \citet{BroeckerSmith2007} and plot the observed relative frequencies for a bin $B_k$ versus $\bar{\pi}_k$. To illustrate the sharpness of the probabilistic forecasts we add histograms for the frequency of usage of the different bins. High-probability forecasts of heavy precipitation amounts were issued very infrequently, and so we plot the inset histograms on a
log-10 scale, providing a better visualization of the distribution in the tails.

Reliability diagrams allow one to graphically assess both reliability and resolution of probabilistic forecasts. A quantitative measure for these characteristics is obtained from the Brier score decomposition \citep{Murphy1973}. We add this information to our diagrams, using a slightly different decomposition proposed by \citet{FerroFricker2012} which is less biased in finite samples and therefore provides a more accurate measure of performance. The observed relative frequencies displayed in the reliability diagram are subject to substantial sampling variability, and so are the estimated REL, RES, and UNC components of the Brier score. To extract an estimate of the uncertainty in these statistics we use bootstrap resampling similar to \citet{Hamill&2007}. Accounting again for dependence of forecast errors in space but assuming independence in time, $90\%$ confidence intervals for the observed relative frequencies on the one hand and REL, RES, and UNC on the other hand were estimated from a 1000-member block 
bootstrap sample \citetext{following \citealp{EfronTibshirani} and \citealp{Hamill1999}} and placed into the reliability diagrams.

\subsection*{Comparison with state of the art post-processing approaches}

We first compare the performance of our EMOS method from Sec.~\ref{sec:2} with that of extended logistic regression (LR) and Bayesian model averaging (BMA). \citet{SchmeitsKok2010} compared the latter two approaches using ECMWF ensemble precipitation re-forecasts over the Netherlands for medium to long forecast ranges, and found that after some modification of BMA both methods performed similar. The present setup with short-range forecasts from a high-resolution NWP model may, however, present rather different challenges and the conclusions concerning the performance of the two methods need not be the same.

Extended LR was proposed by \citet{Wilks2009} and fits a censored logistic distribution with a mean related to the (power-transformed) ensemble mean, to power transforms of observed precipitation amounts. At its core it is an EMOS approach similar to ours, but the model is rewritten such that the model fitting can be done within a logistic regression framework. Each observation $y_s$ is turned into a number of binary responses corresponding to exceedance of certain climatological quantiles specific to site $s$. By including a function $g(q)$ of the respective quantile in the set of predictors, the LR fit can be done for all quantiles simultaneously, and the fitted model corresponds to a full predictive distribution. We implement this method as described in \citet{Wilks2009} with climatological quantiles derived from rain gauge observations between 2005 and 2010. Since the quantiles $q_{0.05}, q_{0.1}, q_{0.33}$ and $q_{0.5}$ are equal to zero for almost all stations, and the other quantiles used in \citet{
Wilks2009} are relatively small, we use $q_{0.98}$ as an additional threshold. After communication with Zied Ben Bouall\`{e}gue from Deutscher Wetterdienst, who uses a very similar forecast setup \citep{BenBouallegue2012} we model square roots of precipitation accumulations, i.e.\ $g(q)=b_2\sqrt{q}$, and use the fourth root of the ensemble mean as a predictor. The latter yields a considerable improvement in our setup compared to the original choice.

BMA was first developed for temperature \citep{Raftery&2005}, a variant for precipitation was proposed by \citet{Sloughter&2007}. Each ensemble member is associated with kernel density $p_k(\tilde{y}_s|f_{sk})$, which for precipitation consists of a discrete (point mass at zero) and a continuous (gamma distribution) component which is fitted to cube root $\tilde{y}$ of observed precipitation amounts. The final predictive density is then a mixture
\[
 p(\tilde{y}_s|f_{1s},\ldots,f_{Ks}) = \sum_{k=1}^K w_k \: p_k(\tilde{y}_s|f_{ks}),
\]
of the members' individual densities with weights $w_1,\ldots,w_K$ that reflect each member's skill. BMA is implemented in the 'ensembleBMA' package in R \citep{Fraley&2011}, and we use it with the control options \verb tol=1e-4 , which stabilizes the estimates of the weights \citep{Hamill2007}, and \verb power=0.5 , which gives a much better fit for our data set than the default cube root transformation.

Extended LR and BMA were both fitted with the same training sets as our EMOS method, consisting of observations of all available stations with identical parameters and the 30 training days preceding every verification day. Brier skill scores for different threshold values and CRPSSs are shown in Table \ref{Tab:SkillScores}. All three methods improve on the probabilistic forecasts derived from the raw ensemble with the performance of extended LR and EMOS being approximately equal and BMA a little bit behind. Is the improvement of our method over the raw ensemble and the reference methods statistically significant or may it simply be a lucky coincidence? To answer this question we tested for equal performance of EMOS and its competitors and obtained the p-values:
\[
  \mbox{ensemble: } <0.01, \quad \mbox{ext.~LR: } 0.63, \quad \mbox{BMA: } <0.01
\]
These values suggest our method attains significantly better CRPSs than the raw ensemble and BMA, while the improvement over extended LR is not significant. A thorough analysis of reliability diagrams (not presented here) and the corresponding Brier score decomposition shows that BMA tends to underpredict precipitation amounts at medium to high levels. Extended LR predictions are more reliable than the ones by EMOS, but the latter have better resolution. This is plausible since our method is based on the untransformed ensemble mean and therefore stays closer to the raw ensemble forecasts. Adjustments of our predictors might well improve reliability and even the overall skill, but we contend that extensions along the lines of Sec.~\ref{sec:3} are a preferable means to improve reliability, and may be able to achieve this without trading such improvement for lower resolution. That this is indeed the case will be shown in the next subsection.

\begin{figure*}
 \centering
 \includegraphics[width=1\textwidth]{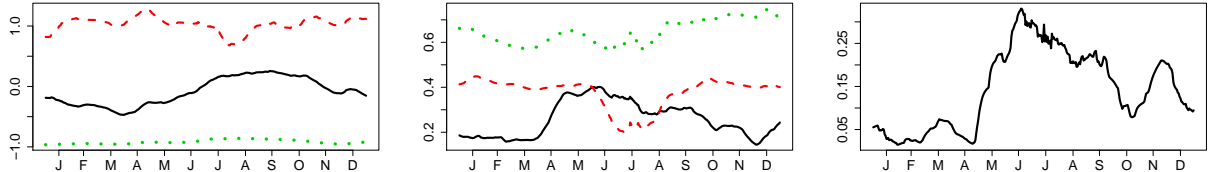}
 \caption{Temporal evolution of the location (left) and scale (middle) parameters, and of the shape parameter $\xi$ (right) for EMOS using a 60 km neighbourhood. $\alpha_i$ and $\beta_i, \: i=1,2,3$ are plotted as black solid lines ($i=1$), red dashed lines ($i=2$) and green dotted lines ($i=3$), respectively.}
 \label{Fig:TempCycleN60}
\end{figure*}

\subsection*{The benefit of using neighbourhood information}

The general idea of considering forecasts from a larger neighbourhood around the location of interest has been discussed previously. \citet{Theis&2005} use deterministic forecasts within a spatio-temporal neighbourhood as a substitute for a forecast ensemble and derive probabilistic statements from it. Similar ideas are used by \citet{BenBouallegue&2012} to enhance the COSMO-DE ensemble. \citet{BentzienFriederichs2012} consider an enhanced time-lagged ensemble and derive from it different predictors for post-processing. Our approach is in the same spirit, but sets itself apart through the distance based weighting scheme, the employment of certain ensemble statistics as predictors of the {\em scale} of the predictive distribution, and in particular through the distinction between ensemble uncertainty and uncertainty due to displacement errors. To analyse the role of these scale predictors we depict the temporal evolution of all parameters of our EMOS method with a 60 km neighbourhood in Fig.~\ref{Fig:TempCycleN60}. The bias parameters are very similar to those obtained without neighbourhood information, and the shape parameter exhibits the same seasonal cycle with a slightly smaller amplitude (unlike in Fig.~\ref{Fig:TempCycleLoc} it is now always positive). More interesting is the role of the two aforementioned predictors for the scale parameter, highlighted by the temporal evolution of $\beta_1$ (ensemble uncertainty) and $\beta_2$ (spatial uncertainty). The latter is constantly higher, suggesting that displacement errors are indeed an important factor in prediction uncertainty. Moreover, the drop of $\beta_1$ together with the increase of $\beta_0$ during the summer months suggests that sources of uncertainty behind the processes causing precipitation in summer are still hard to capture, while spatial uncertainty is present throughout the year.

\begin{figure*}
 \centering
 \includegraphics[width=1\textwidth]{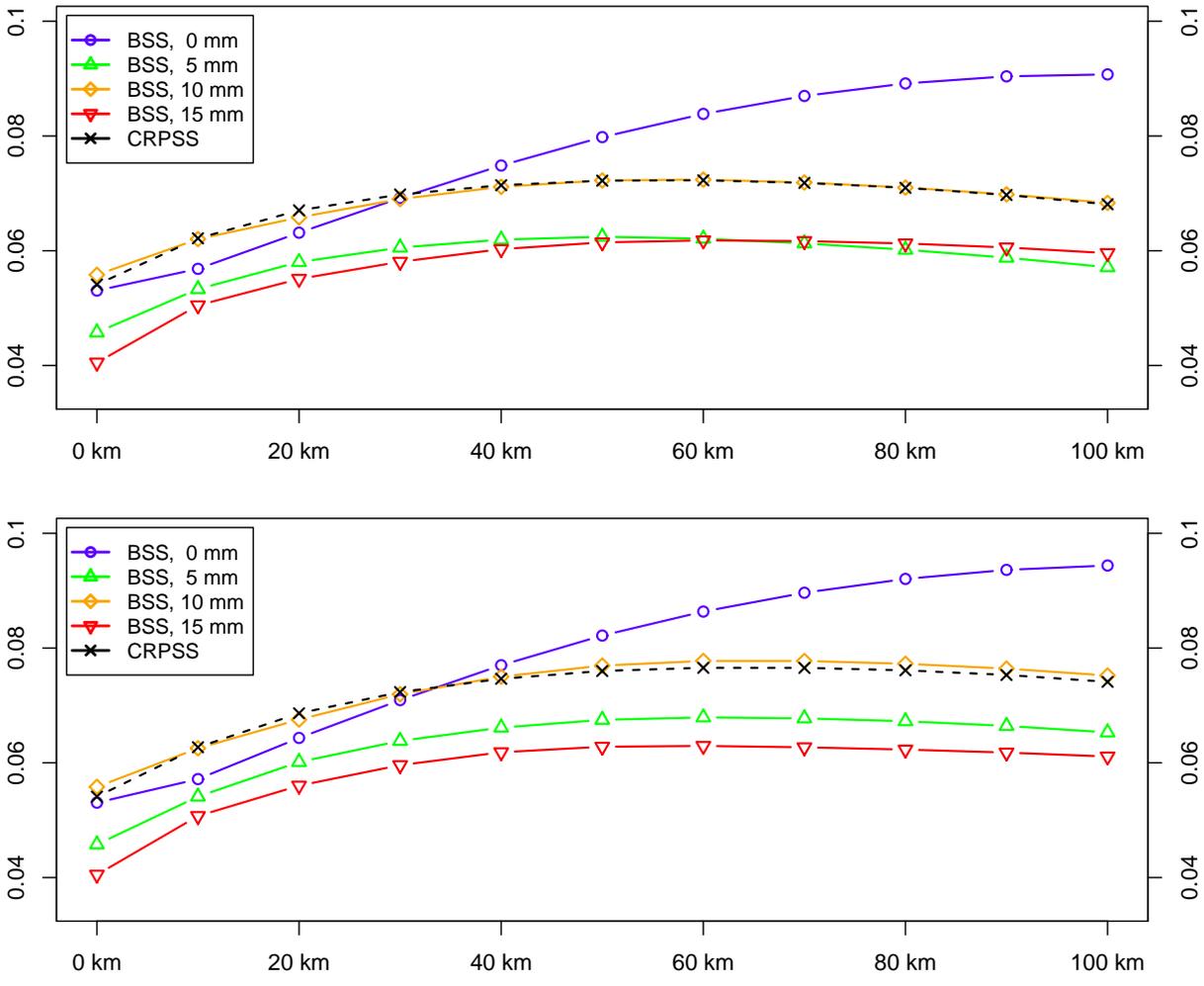}
 \caption{BSS and CRPSS with the EMOS method and different neighbourhood sizes with (bottom) and without (top) climatological correction.}
 \label{Fig:SkillScores}
\end{figure*}

We now study the effect of neighbourhood size on the predictive performance of EMOS forecasts, and check whether the climatological correction suggested in Sec.~\ref{sec:3} yields indeed a notable improvement. Figure \ref{Fig:SkillScores} depicts the CRPSS and the BSS for different neighbourhood sizes. Apparently, the use of neighbourhood information can improve the predictive performance considerably, both with lower and higher thresholds. The biggest improvement is obtained for a neighbourhood radius of 60 km for which the CRPSS is about 0.072. The climatological correction further increases this value to about 0.076. As can be expected, its effect increases with neighbourhood size and it slows the decline of the forecast skill after the maximum at 60 km. An analysis of the score differences calculated separately for each verification date (not presented here) shows that the correction enhances the benefit of using neighbourhood information on the majority of days. However, there are also a few days where 
results would be better without climatological correction, which suggests that not every precipitation event is influenced by orographic effects. Is the improvement in overall predictive performance statistically significant? Table \ref{Tab:p-values} gives p-values for tests for equal performance of EMOS methods using different neighbourhoods. In the first two rows, each value corresponds to a comparison of a neighbourhoods with radius $r$ and $r-10$ km, respectively. The 10 km neighbourhood is compared with the simple EMOS method from Sec.~\ref{sec:2}. For up to 40 km (50 km if climatological correction is used) increasing the radius of the neighbourhood significantly increases the performance, while increasing the neighbourhood radius to more than 80 km (90 km if climatological correction is used) results in in a significant decline of performance. A direct comparison of the results with and without climatological correction shows that the score differences are significant at the $5\%$ level for all 
neighbourhood sizes. Comparing the top and bottom plot in Fig.~\ref{Fig:SkillScores} it may seem surprising that even very small score differences are clearly significant. This can be explained by the fact that the t-test statistic depends not only on the average score difference, but also on the standard deviation of daily differences. The forecasts with and without climatological correction are rather similar with the latter being slightly better. This results in very small standard deviations of daily score differences and thus renders even small score differences significant. Overall we conclude that neighbourhood sizes between 50 km and 80 km yield the most skilful probabilistic forecasts with the maximum being attained for $r=60$ km. This is in agreement with \citet{JohnsonWang2012} who study neighbourhood-based probabilistic forecasts using a 48 km radius and \citet{Mittermaier&2011} who found the high-resolution (4 km) Unified Model (MetUM) to become skilful at a scale of about $30-35$ km (note that 
the effective radius of our approach is smaller than $r$ because gridpoints near the boundaries of the neighbourhoods are assigned very low weights).

\begin{table*}\footnotesize
 \caption{P-values obtained by testing for equal performance of the EMOS method using a neighbourhood with radius $r$ and the same method with neighbourhood radius $r-10$ km. The first two rows correspond to the variants with and without climatological correction, the last row tests, for each radius, if the different between the two variants is significant.}
 \centering
 \label{Tab:p-values}
 \begin{tabular}{lcccccccccc}
  \toprule
   & 10 km & 20 km & 30 km & 40 km & 50 km & 60 km & 70 km & 80 km & 90 km & 100 km \\
  \midrule
  without cl.~corr. & $<0.01$ & $<0.01$ & $<0.01$ & $0.019$ & $0.22$ & $0.90$ & $0.41$ & $0.098$ & $0.013$ &  $<0.01$ \\
  with cl.~corr. & $<0.01$ & $<0.01$ & $<0.01$ & $<0.01$ & $0.015$ & $0.27$ & $0.98$ & $0.35$ & $0.074$ &  $<0.01$ \\
  \midrule
  with vs.~without & $<0.01$ & $<0.01$ & $<0.01$ & $<0.01$ & $<0.01$ & $<0.01$ & $<0.01$ & $<0.01$ & $<0.01$ & $<0.01$\\
  \bottomrule
 \end{tabular}
\end{table*}

We finally take a closer look at the reliability and resolution of our best-performing method (the one with neighbourhood radius 60 km and climatological correction) and compare with the raw ensemble forecasts and with the EMOS method based on the local forecasts only. The confidence intervals in Fig.~\ref{Fig:ReliabilityDiagrams} are rather wide for the high threshold values, indicating that both reliability and resolution vary considerably from day to day. Yet, it is apparent that both EMOS methods strongly improve forecast reliability, even though deviations from the diagonal at the 0 mm and 5 mm level appear to be systematic and might be caused by the lack of flexibility that comes with the assumption of a parametric distribution. This assumption, however, turns out to benefit probabilistic forecasts at high thresholds: 1075 out of 1136 rain gauges report 6-h precipitation accumulations of more than 15 mm on less than 1\% of all days in 2011. Despite the resulting lack of training cases, our method 
yields forecasts that are reliable and have good resolution. Surprisingly, the use of neighbourhood information does not only improve the reliability, but also the resolution of the EMOS predictions. This may be somewhat counter-intuitive since the neighbourhood-based statistics smooth the ensemble forecast and thus flatten the peaks. An explanation for this might be, that the improvement of reliability for many post-processing methods often goes along with pulling the forecasts a bit towards climatology. If the reliability is partly improved by moderately smoothing the forecasts, the subsequent post-processing correction might be more moderate, so that the resolution loss due to smoothing is eventually offset.

\begin{figure*}
 \centering
 \includegraphics[width=\textwidth]{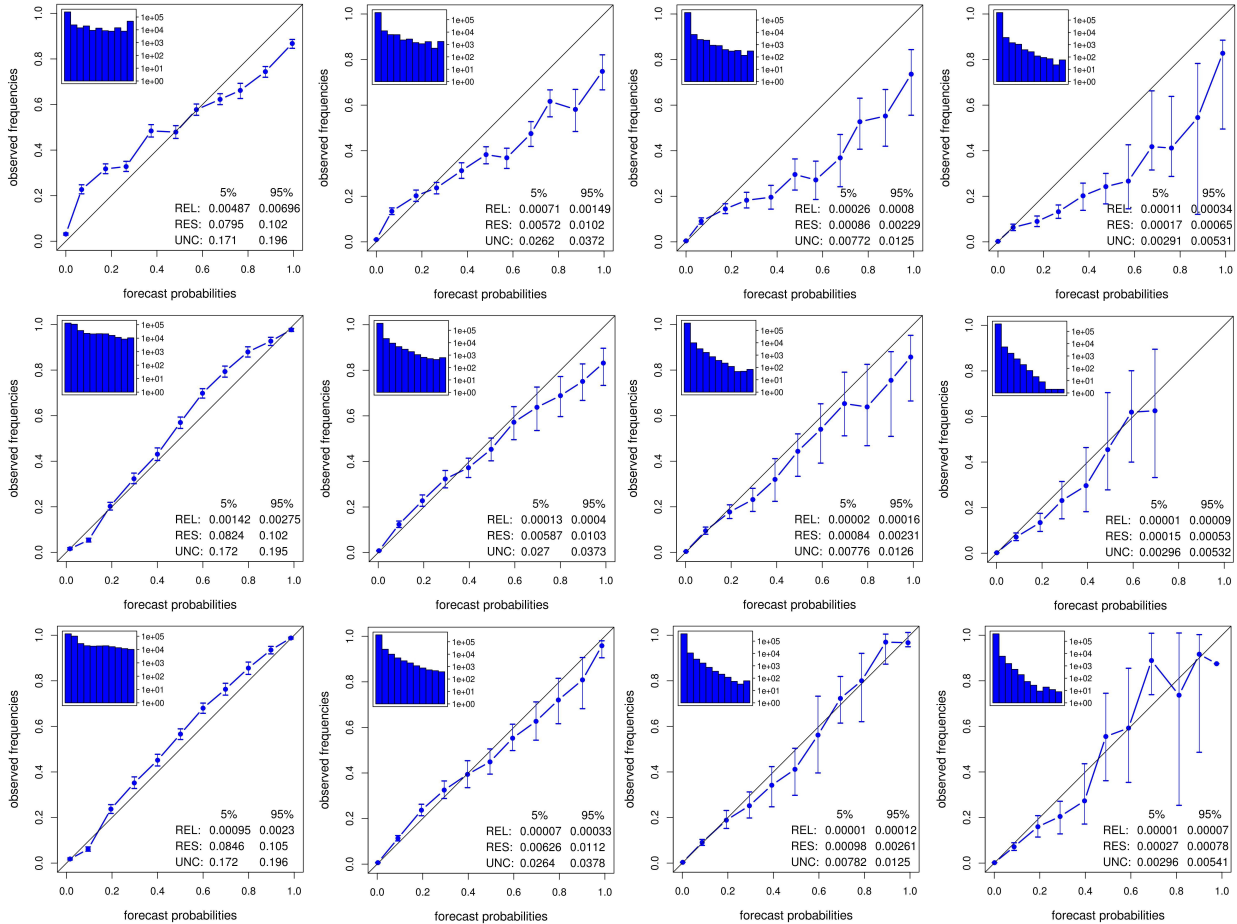}
 \caption{Reliability diagrams for raw ensemble forecasts (top) and probabilistic forecasts by EMOS based on information only (middle) and using a 60 km neighbourhood and climatological correction (bottom). From left to right we consider exceedance of 0.1 mm, 5 mm, 10 mm, and 15 mm precipitation in 6h. The insets show histograms of the log-frequency of cases within the respective bins, the bars correspond to 90\% confidence intervals obtained by bootstrap resampling, and in the bottom right corner 90\% confidence intervals for reliability, resolution, and uncertainty are given.}
 \label{Fig:ReliabilityDiagrams}
\end{figure*}

\section{Discussion}
\label{sec:5}

The EMOS method presented in Sections \ref{sec:2} and \ref{sec:3} was shown to be effective for calibrating precipitation forecasts by the COSMO-DE-EPS. It improved reliability while maintaining the good resolution of the raw ensemble at threshold values up to 15 mm (6 h)$^{-1}$. It still needs to be tested if the model also performs well with even more extreme thresholds, but a much larger verification data set would be required in order to come to authoritative conclusions. It might well turn out that further statistics of the ensemble forecasts in the neighbourhood (e.g.~high quantiles) must be incorporated in the predictive distribution to convey information on heavy precipitation events. In the settings considered in this paper, even our basic method compared well with extended LR and BMA, two state-of-the-art post-processing approaches, and the advanced version using neighbourhood information yielded further improvement. We presume that our approach works well also for other ensemble prediction systems.
\\
There are several lines along which our method could still be extended and improved. A contentious issue is certainly the fact that we use the same post-processing parameters over the whole forecast domain. We do so in order to gather sufficient training data to get stable estimates without having to rely on several years of forecast records (which are not available in our case). For forecast domains with complex topography or consisting of sub-domains with very diverse precipitation climatologies it may be more appropriate, however, to let the parameters vary spatially. This could be realized as an extension of our method analogously to \citet{Kleiber&b2011} where post-processing parameters are fitted locally and geostatistical methods are used to extrapolate them to locations where observations for model fitting are not available. An alternative strategy for locally adaptive post-processing could be the inclusion of further predictors such as orographic elevation or orographic slope (or suitable transforms 
of them) with explanatory power for spatial variations in post-processing.\\
A further simplification in our approach is that it does not account for potentially different skill of the ensemble members. This could be done as in \citet{Gneiting&2005} by using the individual ensemble forecasts rather than the ensemble mean as predictors for the location parameter $m$. We have refrained from doing this here because it would increase the number of model parameters from $7$ to $26$ and entail the danger of overfitting. However, while COSMO-DE-EPS members have been constructed such that they all have comparable skill \citep{Gebhardt&2011}, overfitting may be a price worth paying if one uses an EPS where some members are distinctly less skilled than others. A completely different weighting strategy that we plan to investigate in the future could try to make use of the insights by \citet{KeilCraig2011} who found that both spread and skill of the different subgroups of members in COSMO-DE-EPS depend on the meteorological situation, i.e.\ on whether precipitation is stratiform, due to 
equilibrium or due to non-equilibrium convection. One might even go one step further and check if the seasonal cycle of our post-processing parameters (see Fig.~\ref{Fig:TempCycleN60}) can actually be attributed to seasonal variations in the frequency of weather regimes.  In this case further improvement of the reliability of probabilistic forecasts might be possible if training days are selected by similarity of meteorological situations rather than temporal proximity. In a slightly different setting this has been done successfully by \citet{Kober&2012-1} and their ideas might be transferable to our framework.\\
Finally, there is the issue of spatial consistency, which becomes important when the interest is in forecasting spatially aggregated quantities like the overall precipitation amount in some river catchment, rather than precipitation amounts at individual sites. The EMOS method presented so far only yields univariate predictive distributions, but approaches similar to the one by \citet{Berrocal&2008} could be used to extend it to a multivariate distribution model which can be used to simulate calibrated and spatially consistent precipitation fields. \citet{Sigrist&2012} present a spatio-temporal model that accounts for dependencies between both multiple locations and multiple look-ahead times, and can model phenomena such as transport and diffusion. A combination of their ideas with the approach presented here may be able to attain the ultimate goal of producing calibrated, sharp, and physically realistic probabilistic forecasts of spatio-temporal precipitation fields, but quite some effort still seems to be 
required to realize this in such a way that the resulting method is suitable for operational use. A much simpler approach which could be used right away and may also yield good results is described in \citet{SchefzikDA}: ensemble copula coupling (ECC) can be combined with any univariate post-processing method and transfers the dependence structure of the raw ensemble to samples from the univariate predictive distribution. The EMOS method for precipitation presented in this article complements existing methods for temperature \citep{Gneiting&2005}, wind speed \citep{ThorarinsdottirGneiting2010}, wind vectors \citep{Schuhen&2012} and wind gusts \citep{ThorarinsdottirJohnson2012}. It is reasonably simple and can serve as a basis for future developments along the lines described above.

\section*{Acknowledgement}
The author is grateful to Zied Ben Bouall\`{e}gue, Petra Friederichs, Tilmann Gneiting, and Thordis Linda Thorarinsdottir for helpful discussions.\\
He thanks Sabrina Bentzien and all members of the COSMO-DE-EPS team of Deutscher Wetterdienst (DWD) for their support with the acquisition of the ensemble forecast data. Martin G\"ober kindly provided the rain gauge dataset from the DWD network. Additional rain gauge data (annual total precipitation amounts between 2005 and 2010) were provided by MeteoSwiss, Federal Office of Meteorology and Climatology in Switzerland, and ZAMG, Central Institute for Meteorology and Geodynamics in Austria.\\
The reviewers' comments on an earlier version of this manuscript have helped to improve it significantly.\\
This work is funded by Deutscher Wetterdienst in Offenbach, Germany, in the framework of the extramural research program.

\bibliographystyle{plainnat}
\bibliography{../QJRMS/MSbib}

\end{document}